\documentclass[conference,10pt]{IEEEtran}
\IEEEoverridecommandlockouts
\usepackage{amsmath,amssymb,amsfonts}
\usepackage{algorithmic}
\usepackage{graphicx}
\usepackage{textcomp}
\usepackage{xcolor}
\usepackage{subcaption}
\def\BibTeX{{\rm B\kern-.05em{\sc i\kern-.025em b}\kern-.08em
    T\kern-.1667em\lower.7ex\hbox{E}\kern-.125emX}}

\usepackage{comment}
\usepackage[backend=biber,sorting=none]{biblatex}
\usepackage{multirow}
\usepackage{tikz}
\usetikzlibrary{arrows.meta,positioning,fit,calc,backgrounds}

\begin{document}

%\title{SCARP: Scalable Congestion-Aware Replica Placement for LLM Serving on Heterogeneous Edge Fleets}
\title{Scalable Placement for Disaggregated LLM Serving on Heterogeneous Edge Fleets}

%\begin{comment}
\author{
    \IEEEauthorblockN{Truong-Thanh Le\IEEEauthorrefmark{2}, 
                      Hoang-Loc La\IEEEauthorrefmark{4}, 
                     Amir Taherkordi\IEEEauthorrefmark{2},
                     Frank Eliassen\IEEEauthorrefmark{2} and
                     Phuong Hoai Ha\IEEEauthorrefmark{4}} 

    \begin{tabular}[t]{@{}c@{}}
    \IEEEauthorblockA{\IEEEauthorrefmark{2} 
                      Department of Informatics \\
                      \textit{University of Oslo}\\
                      Oslo, Norway \\
                      \{truongl,amirhost,frank\}@ifi.uio.no} 
    \end{tabular}\nobreak\qquad

    \begin{tabular}[t]{@{}c@{}}
    \IEEEauthorblockA{\IEEEauthorrefmark{4} 
                      Department of Computer Science \\
                      \textit{UiT The Arctic University of Norway}\\
                      Tromsø, Norway \\
                      \{hoang.l.la,phuong.hoai.ha\}@uit.no}
                      \end{tabular}   
}
%\end{comment}

%\author{\IEEEauthorblockN{Anonymous Author(s)}
%\IEEEauthorblockA{Paper submitted for double-blind review}}

\maketitle

\begin{abstract}
Serving large language models (LLMs) at the edge is increasingly desirable for privacy-sensitive applications and scenarios that require data to remain on local devices. However, edge environments are constrained by limited computational resources and exhibit substantial device heterogeneity, making efficient LLM serving challenging. These limitations become particularly pronounced under bursty demand, where congestion and queue buildup can significantly increase response latency. Replicating the model into multiple pipeline instances can improve serving capacity and reduce congestion. Yet, deciding how devices should be grouped into replicas, assigning replicas to prefill or decode functions, and partitioning model layers across heterogeneous devices jointly creates a large combinatorial optimization problem. We present \textsc{Spade}, a scalable and runtime-efficient placement planner for disaggregated LLM serving with pipeline replication across heterogeneous edge clusters. \textsc{Spade} introduces a dominance-based pruning technique that discards inferior replica candidates and recursively eliminates larger candidates that contain them, dramatically shrinking the search space. It further combines a water-filling approach for layer partitioning with a bounded multidimensional knapsack formulation that optimizes placement under a congestion-aware objective. On a synthetic fleet of 30 devices spanning five hardware classes, \textsc{Spade} finds the optimal placement in 441\,ms. In contrast, a genetic-search baseline augmented with dynamic-programming partitioning requires approximately two hours and still produces solutions with a 19.7\% worse congestion objective. When evaluated on fleets containing ten device types, \textsc{Spade}'s pruning strategy eliminates 99.6\% of candidate placements while yielding the same surviving candidate set as exhaustive enumeration.
\end{abstract}


\section{Introduction}\label{sec.intro}
Large language models (LLMs) are increasingly moving from centralized datacenters to edge deployments \cite{chen2026network}, driven by privacy requirements that prevent organizations and individual users from sending sensitive data to third-party APIs \cite{yan2025protecting}.
%Rather than running on homogeneous clusters, these deployments must operate on the available hardware \cite{li2026prima,tong2025parallax,sun2025collaborative} 
Instead of running on homogeneous clusters, these deployments must operate on the available hardware \cite{li2026prima,tong2025parallax,sun2025collaborative} 
%desktops equipped with NVIDIA GPUs, laptops powered by AMD accelerators, Macs with unified memory, and embedded devices 
connected through commodity local-area networks rather than high-speed interconnects such as NVLink. Hence, the execution environment is inherently heterogeneous in architecture, memory capacity, and communication bandwidth \cite{mei2025helix,zhao2024poster,mei2026coral}. 

Deploying modern LLMs on such fleets is challenging because model sizes, activations and KV cache usage often exceed the memory capacity of any single device \cite{li2026prima,sun2025collaborative}. Memory therefore is one of the main factors that determines both where the model can be placed and how many requests can be served concurrently \cite{le2026e2llm}. 
%Consequently, \textit{pipeline parallelism} is not merely a performance optimization in heterogeneous edge fleets; it is the fundamental mechanism that enables LLM serving in the first place \cite{hu2022pipeedge,zhang2024edgeshard,le2024optimal}.
Consequently, \textit{pipeline parallelism} is the fundamental mechanism that enables LLM serving in the first place \cite{hu2022pipeedge,zhang2024edgeshard,le2024optimal}.
Moreover, LLM serving is itself heterogeneous. \textit{Prefill} is compute-intensive and processes an entire prompt at once, whereas \textit{Decode} is memory-intensive and generates tokens iteratively. This heterogeneity motivates LLM disaggregation in datacenter systems \cite{patel2024splitwise,zhong2024distserve}. At the edge, disaggregation becomes even more challenging because the available devices differ substantially.% in compute performance and memory capacity.
%Moreover, LLM serving is itself heterogeneous. \textit{Prefill} is compute-intensive and processes an entire prompt at once, whereas \textit{decode} is memory-intensive and generates tokens iteratively. This motivates LLM disaggregation in datacenter systems \cite{patel2024splitwise,zhong2024distserve}, and becomes even more complicated at the edge, where devices differ substantially.% in exactly these two resources.

An edge fleet must also serve many concurrent requests with limited compute and memory capacity, so requests can readily arrive faster than it can process them; time-to-first-token (TTFT) then becomes dominated by congestion, the time requests spend waiting in queues~\cite{arun2026towards}. 
%Deploying multiple \emph{pipeline replicas}, each a complete copy of the LLM partitioned across a subset of devices, alleviates congestion but turns deployment into a placement problem: devices must be allocated across replicas, each replica assigned to Prefill or Decode, and layers partitioned across each replica's devices. 
Deploying multiple \emph{pipeline replicas}, each a complete copy of the LLM deployed across a subset of devices, alleviates congestion but turns deployment into a placement problem: devices must be allocated across model replicas, each model replica assigned to Prefill or Decode, and layers partitioned across each replica's devices.
Existing solutions address only parts of this problem. 
Halda~\cite{li2026prima} and LOIP~\cite{sun2025collaborative} optimize single-pipeline deployments for single-request latency, while Helix~\cite{mei2025helix} and Parallax~\cite{tong2025parallax} maximize aggregate throughput without Prefill--Decode role assignment. E2LLM~\cite{le2026e2llm} is closest to our work, targeting congestion with Prefill and Decode replicas; however, its objective combines Prefill service time with congestion, which can favor Prefill capacity even when it does not shorten queues, and its genetic search provides no optimality guarantee. What remains missing is an objective that isolates congestion, together with a search that finds the optimal placement at fleet scale.

Finding this optimum is hard because placement nests two problems. The outer problem allocates devices to replicas and assigns each replica a phase; it is NP-hard~\cite{jiang2025hexgen2,jiang2025thunderserve}, with up to $2^N$ candidate replicas for $N$ devices. The inner problem partitions the model's layers across a candidate's devices to maximize throughput under memory constraints; existing solutions solve it by dynamic programming (DP) over an ordered device sequence~\cite{le2026e2llm,hu2022pipeedge,zhang2024edgeshard} in $O(KL^2)$ time for $K$ devices and $L$ layers. Because a candidate's throughput is unknown until its inner problem is solved, the exact optimum requires an inner solver for exponentially many candidates: on a 30-device fleet with ten device types, exhaustive evaluation takes over nine hours even with a fast inner solver, and this cost grows exponentially as fleets grow. 
%Planning time is thus the obstacle to optimality, and existing planners keep it tractable either through heuristic search, namely tabu search, evolutionary search, simulated annealing, and genetic algorithms in ThunderServe~\cite{jiang2025thunderserve}, HexGen~\cite{jiang2024hexgen}, HexGen-3~\cite{jiang2026hexgen3}, and E2LLM~\cite{le2026e2llm}, respectively, or by restricting the problem: Parallax~\cite{tong2025parallax} decomposes placement into sequential stages, and the exact formulations of Helix~\cite{mei2025helix} and Halda~\cite{li2026prima} optimize layer placement rather than replica formation.
Planning time is thus the obstacle to optimality, and existing planners address it through heuristic search with no optimality guarantee, namely tabu search~\cite{jiang2025thunderserve}, evolutionary search~\cite{jiang2024hexgen}, simulated annealing~\cite{jiang2026hexgen3}, and genetic algorithms (GA)~\cite{le2026e2llm}, or by restricting the problem: Parallax~\cite{tong2025parallax} decomposes placement into sequential stages, while the exact formulations of Helix~\cite{mei2025helix} and Halda~\cite{li2026prima} solve layer placement (i.e., the inner problem) rather than replica formation. Using these inner solvers would also not reduce planning time either, since each requires solving an integer program for every candidate, which is still more costly than DP.

Our new solution presented in this paper, \textsc{Spade}, avoids both compromises. It solves the outer problem exactly in time exponential only in the number of device types, which is typically small in heterogeneous edge fleets, rather than in the number of devices. 
%Exploiting the interchangeability of devices of the same type, Spade represents a candidate replica by its per-type device counts and formulates placement as a bounded multidimensional knapsack, solved exactly by DP or, when memory is limited, by a mixed-integer program.
Exploiting the interchangeability of devices of the same type, Spade represents a candidate replica by its per-type device counts and formulates allocation over the generated candidates as a bounded multidimensional knapsack, solved exactly by DP or, when memory is limited, by a mixed-integer program.
%To further reduce planning time, Spade introduces a dominance rule that discards any candidate whose devices can be decomposed into smaller candidates with equal or greater aggregate throughput, and skips generating larger candidates that contain it. Because splitting a replica duplicates the model weights, a decomposition can outperform its parent only when throughput is limited by pipeline latency rather than memory, a limitation that larger candidates inherit; the latter step relies on an assumption that held for every instance we verified exhaustively.
\textsc{Spade} also reduces planning time by introducing a \textit{dominance rule} that discards any candidate whose devices can be decomposed into smaller candidates with equal or greater aggregate throughput, and skips generating larger candidates that contain it. Since splitting duplicates model weights, a decomposition can outperform its parent only when throughput is latency-bound rather than memory-bound. 
%We assume this limitation is inherited by larger candidates containing the parent; this assumption held in every instance we verified exhaustively.
We assume that larger candidates containing the parent inherit this limitation and can therefore be skipped. Across all instances that we could enumerate exhaustively (Table III), this assumption eliminated no optimal candidate.
Additionally, because only the head and tail of a replica's pipeline are distinct \cite{le2026e2llm}, the inner problem admits a closed-form water-filling solution that reduces complexity from $O(KL^2)$ to $O(K)$. 
%Finally, Spade models disaggregated serving as a fluid tandem queue and introduces an objective that isolates congestion from service time, minimizing congestion while ensuring that every replica sustains a minimum per-request token rate.
%In summary, we make the following contributions: 
In summary, our contributions are:
\begin{itemize} 
    \item A congestion-aware objective for disaggregated LLM serving that models the system as a fluid tandem queue and isolates congestion from Prefill service time (Section~\ref{subsec:method:objective}). On an eight-device CUDA--Metal testbed, placements selected by this objective reduce worst-case waiting time by up to 2.44$\times$ compared with the state-of-the-art objectives (Section~\ref{subsec:eval:objective}). 
    \item A scalable placement solution that combines a type-count knapsack formulation, dominance pruning, and closed-form water-filling partitioning (Sections~\ref{subsec:method:inner}--\ref{subsec:method:outer}). On a synthetic 30-device fleet, dominance pruning evaluates up to 99.6\% fewer candidates than exhaustive enumeration, and Spade reaches the verified optimum in 441~ms, whereas the state-of-the-art method requires 6855~s and yields 19.7\% higher congestion (Section~\ref{subsec:eval:planner}). 
    \item An open-source peer-to-peer serving framework built on \texttt{llama.cpp}. Unlike the star topology used by \texttt{llama.cpp} and E2LLM, which routes all activations through a single node, the framework enables fully decentralized communication (Section~\ref{subsec:eval:framework}).% and underlies all experiments in this paper.
\end{itemize}

\section{Related Work}\label{sec.relatedwork}
\subsection{Distributed LLM inference on heterogeneous devices}
\label{subsec:rw:placement}

Partitioning a model across multiple devices is well established. PipeEdge and EdgeShard partition transformer layers under memory constraints \cite{hu2022pipeedge,zhang2024edgeshard}; LOIP, Parallax, and Helix extend placement to heterogeneous deployments with communication, scheduling, and optimization constraints \cite{sun2025collaborative,tong2025parallax,mei2025helix}; and prima.cpp (Halda) targets consumer hardware through integer-programming placement and disk offloading \cite{li2026prima}. LOIP and prima.cpp optimize single-pipeline deployments for single-request metrics such as TPOT and latency, whereas Parallax and Helix maximize aggregate throughput. However, none of these systems models congestion explicitly as a placement objective or assigns distinct Prefill and Decode roles to different replicas.
%While E2LLM permits a heterogeneous fleet to be divided into multiple replicas \cite{le2026e2llm}, placement decisions remain driven by service-time objectives rather than queueing behavior. 
%While E2LLM instantiates multiple LLM replicas by assigning different subsets of devices to different model copies \cite{le2026e2llm}, placement decisions remain driven by service-time objectives rather than queueing behavior. 
%We additionally find that prima.cpp's multi-round SSD--CPU--GPU pipeline is effective only in its CPU execution; GPU execution falls back to a single round with one microbatch, eliminating the pipeline and overlap that motivate the design (Section~\ref{sec.eval}). Overall, although existing work accounts for heterogeneity in compute, memory, and communication, placement remains largely disconnected from the congestion that emerges under load.
While E2LLM instantiates multiple Prefill and Decode replicas on heterogeneous edge fleets \cite{le2026e2llm}, its objective couples congestion with Prefill service time.
We additionally find that prima.cpp's multi-round SSD--CPU--GPU pipeline is effective only in its CPU execution; GPU execution falls back to a single round with one microbatch, eliminating the pipeline and overlap that motivate the design (Section~\ref{sec.eval}).

\subsection{Prefill--decode disaggregation}
\label{subsec:rw:disagg}

%Separating the two inference phases onto distinct resources is now established practice. Splitwise assigns prefill and decode to different machine pools, transferring the KV cache between them, but relies on a manually configured pool assignment \cite{patel2024splitwise}. DistServe instead co-optimizes the placement and parallelism of each phase against per-phase latency targets \cite{zhong2024distserve}. HexGen-2 extends disaggregation to heterogeneous GPU clusters, jointly optimizing device grouping, per-group parallel strategies, and KV-cache communication paths \cite{jiang2025hexgen2}. More recent systems refine how work is routed between the two pools: prefill deflection allows decode nodes to absorb chunked prefill steps when doing so improves tail TTFT \cite{arun2026towards}, while SLO-aware allocation analytically sizes the prefill and decode pools under a time-to-first-token target \cite{li2026slo}.

%Our work adopts prefill-decode disaggregation rather than introducing it. The distinction lies in the deployment environment. Existing systems assume homogeneous accelerators, making pool assignment primarily a capacity-planning decision. In heterogeneous edge fleets, however, devices differ substantially in compute performance and memory capacity, precisely the resources stressed by prefill and decode, respectively. Consequently, assigning a device to a prefill or decode role becomes part of the placement problem itself rather than a provisioning decision layered on top of it.

Prefill--decode disaggregation is now a common practice in LLM serving. Splitwise separates the two phases onto dedicated machine pools \cite{patel2024splitwise}, DistServe jointly optimizes their placement and parallelism \cite{zhong2024distserve}, and HexGen-2 extends the approach to heterogeneous GPU clusters \cite{jiang2025hexgen2}. More recent work refines resource allocation through workload-aware routing and analytical provisioning \cite{arun2026towards,li2026slo}. Hence, we adopt disaggregation rather than introducing it. The key difference is that existing systems treat role assignment largely as a provisioning decision, whereas heterogeneous edge fleets contain devices that differ substantially in compute performance and memory capacity, precisely the resources stressed by Prefill and Decode, respectively. Consequently, assigning a device to a phase becomes part of the placement problem itself.

\subsection{Queueing and contention in LLM serving}
\label{subsec:rw:queueing}

As LLM serving systems matured, attention shifted from execution latency to the congestion that emerges under load. Analytical models provision Prefill and Decode capacity under TTFT targets \cite{li2026slo}, while measurement studies show that queueing delay and KV-cache transfer dominate tail TTFT in loaded clusters \cite{arun2026towards}. Subsequent work addresses congestion through scheduling, including Peace \cite{zhang2026peace} and APEX \cite{fan2026apex}, but treats device heterogeneity as largely orthogonal to the optimization. E2LLM extends placement optimization to heterogeneous edge fleets and allows multiple replicas \cite{le2026e2llm}, yet estimates latency by combining Prefill service time with bottleneck-phase congestion. As discussed in Section~\ref{sec.intro}, this couples service costs with queueing effects and can reward additional Prefill capacity without reducing congestion. Our work starts from the same edge-serving setting but models Prefill and Decode as a fluid tandem queue, isolating congestion from service costs and optimizing placement directly around the resulting queueing behavior.

\begin{figure*}[t]
  \centering
  %\resizebox{0.98\linewidth}{!}{\input{texFiles/fig-method}}
  \includegraphics[width=0.9\linewidth]{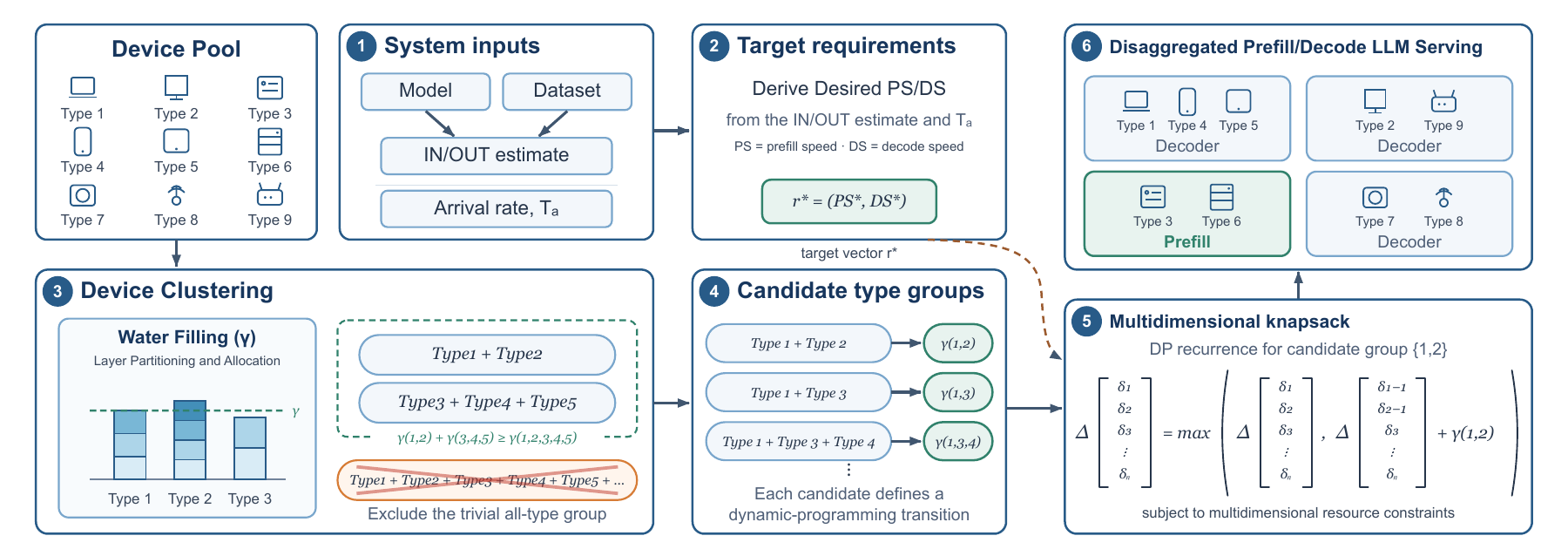}
  \vspace{-0.1cm}
  \caption{\textsc{Spade} system overview. $\gamma$: overall throughput after using layer partitioning. $\delta_i$: number of devices in type $i$}
  \label{fig:Spadesystem}
  \vspace{-0.3cm}
\end{figure*}

\subsection{Multi-cluster placement problem}
\label{subsec:rw:search}

%Once a fleet can be divided into multiple replicas/clusters, placement becomes a combinatorial problem.
Once multiple LLM replicas can be deployed simultaneously, placement becomes a combinatorial problem because devices must be allocated across replicas.
Selecting which devices form a replica is a clustering problem with an exponential search space, and ThunderServe explicitly identifies the problem as NP-hard \cite{jiang2025thunderserve}. Existing systems therefore rely on heuristic search to navigate that complexity. HexGen forms asymmetric groups through evolutionary search \cite{jiang2024hexgen}; HexGen-2 \cite{jiang2025hexgen2} replaces evolutionary search with spectral partitioning refined by Kernighan--Lin, evaluates each candidate through a max-flow formulation that selects per-group parallel strategies and KV-cache communication paths, and further improves solutions through max-flow-guided edge swaps. HexGen-3 employs simulated annealing, a metaheuristic optimization algorithm, to search for a deployment plan \cite{jiang2026hexgen3}. ThunderServe adopts tabu search \cite{jiang2025thunderserve}, while E2LLM uses a GA \cite{le2026e2llm}. These methods substantially reduce search cost and scale to larger deployments, but they offer no optimality guarantee.% that the placement found is optimal. 

Exact optimization has also been explored. Helix formulates placement as a max-flow mixed-integer linear program \cite{mei2025helix}, while Halda decomposes an integer program over layer assignments \cite{li2026prima}. However, these methods optimize layer placement within a pipeline rather than the higher-level problem of selecting and composing multiple replicas from a heterogeneous fleet. Additionally, the challenge is not expressing the optimization problem, but evaluating the vast number of candidate replicas it generates. Before a replica can be compared with another, its throughput must first be estimated by solving an inner layer-partitioning problem. Consequently, the cost of replica evaluation quickly dominates planning time (see Section \ref{subsec:eval:planner}). Reducing this cost is therefore key to the \textit{feasibility and scalability} of the planner. Existing work therefore tends to treat exactness and tractability as competing objectives. Our approach reconciles the two for the outer problem. The dominance rule introduced in Section~\ref{sec.method} eliminates the need to evaluate the overwhelming majority of candidates, removing the principal computational bottleneck and making placement optimization tractable at fleet scale.

\section{Proposed Method}\label{sec.method}
Figure~\ref{fig:Spadesystem} provides an overview of \textsc{Spade}. First, \textsc{Spade} constructs list of candidates for the outer problem and solves inner problem with water-filling formulation (Section \ref{subsec:method:inner}). Additionally, \textsc{Spade} applies the dominance rule while constructing the list to eliminate unpromising candidates and thus reducing the search space (Section \ref{subsec:method:dominance}). Finally, it solves the outer problem (Section \ref{subsec:method:outer}) based on the congestion model of disaggregated Prefill/Decode to determine the placement that minimizes the modelled congestion (Section \ref{subsec:method:objective}).

\subsection{Congestion model for prefill/decode placement}
\label{subsec:method:objective}

%We model a disaggregated deployment consisting of one or more Prefill replicas?? and one or more Decoder replicas.
We model a disaggregated deployment consisting of at least one Prefill replica and one Decoder replica.
Every request must first traverse a Prefill before being forwarded to a Decoder, making the deployment a two-stage tandem queue rather than two independent services. Let $PS$ and $DS$ denote the aggregate Prefill and Decoder throughput of the deployment in tokens per second, $N_P$ the average number of input tokens per request, $N_D$ the average number of generated tokens, and $T_a$ the average inter-arrival period between requests. The aggregate throughputs follow directly from the partition plan described in Section~\ref{subsec:method:inner}, summed over all replicas assigned to each phase. We define
\begin{equation} 
    t_P = N_P/PS, \qquad t_D = N_D/DS%\frac{N_P}{PS},
    %\qquad 
    %t_D = \frac{N_D}{DS}, 
\end{equation}
as the service times of the Prefill and Decoder stages, respectively.
Congestion appears in the Prefill stage ($C_P$) only when requests arrive faster than the stage can serve them, i.e., when the prefill time $t_P$ exceed the average inter-arrival period.
\begin{equation} 
    C_P = \max\!\left(0,\; t_P - T_a\right). 
    \label{eq:cp} 
\end{equation}

The Decoder stage experiences a different arrival process. Once Prefill becomes saturated, requests can leave the first stage no faster than the Prefill service rate. Consequently, the Decoder does not observe the external arrival period $T_a$ directly. Instead, requests arrive at intervals
\begin{equation} 
    T_D = \max\!\left(T_a,\; t_P\right), 
    \label{eq:td} 
\end{equation}
which reflects either the original request stream or the throttled output of a saturated Prefill stage. Measuring Decoder congestion against $T_a$ would therefore count the delay twice. Decoder congestion $C_D$ is instead determined by the excess of its service time over the period at which work actually arrives:
\begin{equation} 
    C_D = \max\!\left(0,\; t_D - T_D\right). 
    \label{eq:cd} 
\end{equation}

Once a queue forms, the delay experienced by later requests accumulates. Each arriving request inherits the backlog generated by earlier requests, causing congestion to grow linearly over the burst. 
\begin{comment}
For a burst of $n$ requests, the total accumulated waiting time is therefore
\vspace{-0.1cm}
\begin{equation} 
    \vspace{-0.1cm}
    W = \left(C_P + C_D\right)n(n-1)/2 
    \label{eq:objective} 
\end{equation}
\end{comment}
\textsc{Spade} selects the placement that minimizes $W=C_P+C_D$. 
%Especially, 
The congestion can happen in either a single phase or both phases. For the former scenarios, allocating more resources on the congested phase would help. For the latter scenarios, 
from Eqs.~\ref{eq:cp}--\ref{eq:cd}, the congestion happens on both phases when $t_D > t_P > T_a$. %In this regime, Decoder is the dominant bottleneck??: 
In this regime, Decoder is the dominant bottleneck: increasing Decoder throughput reduces both Decoder congestion and total waiting time, regardless of congestion in the Prefill stage. Hence, allocating more resources to Decoder is generally more beneficial than increasing Prefill capacity. % even when there is congestion in Prefill.
%In contrast, when congestion happens only on single phase, 

\begin{comment}
The resulting objective has two useful properties. First, it measures congestion rather than execution time. When
\begin{equation} 
    \max(t_P,t_D) \le T_a, 
\end{equation}
the deployment is under-loaded, no queue forms, and therefore $W=0$ regardless of the service time of either phase. Prior formulations add the Prefill service time $t_P$ to the objective. Because $t_P=N_P/PS$ depends on the placement itself, this addition is not a constant offset across candidates; instead, it changes their relative ordering and can favor placements that increase Prefill capacity without reducing congestion.

Second, simultaneous congestion in both phases occurs only in a single operating regime. From Eq.~\ref{eq:cd}, the condition $C_D>0$ requires
\begin{equation} 
    t_D > \max(T_a,t_P), 
\end{equation}
which immediately implies $t_D>t_P$. Both congestion terms are therefore strictly positive if and only if
\begin{equation} 
    t_D > t_P > T_a. 
\end{equation}
In other words, Prefill and Decoder accumulate backlog simultaneously exactly when Decoder is the bottleneck and Prefill is already saturated. Outside this regime, the objective is governed by whichever phase first exceeds the arrival rate, providing a simple regime test at no additional computational cost.
\end{comment}

\subsection{Model partitioning with water-filling}
\label{subsec:method:inner}

Before a placement can be evaluated, we must determine how the transformer's layers are distributed across each replica and estimate the throughput that replica can sustain. 
%A replica consists of a set of devices that collectively host all $L$ transformer layers and execute them as a pipeline-parallel ring. 
A pipeline replica is a complete copy of the LLM deployed across a subset of devices. The assigned devices collectively host all $L$ transformer layers and execute them as a pipeline-parallel ring.
Device $i$ stores a contiguous block of $\ell_i$ layers:
\vspace{-0.1cm}
\begin{equation} 
    \vspace{-0.1cm}
    \sum_i \ell_i = L, 
    \qquad 
    \ell_i \ge \ell_{\min}, 
\end{equation} 
which limits devices assigned to a replica to at most 
\vspace{-0.1cm}
\begin{equation} 
    \vspace{-0.1cm}
    K_{\max} = \left\lfloor \frac{L}{\ell_{\min}} \right\rfloor 
\end{equation} 

Unlike a centralized deployment, positions within a pipeline sequence are not interchangeable. The head device hosts the embedding table and the tail device the final normalization and output projection, making both the fixed memory overhead $c_i(r)$ and fixed stage latency cost $\alpha_i(r)$ depend on the role $r\in\{\text{head},\text{mid},\text{tail}\}$. A replica of $K$ devices therefore admits $K(K-1)$ distinct head--tail assignments \cite{le2026e2llm}%??
, each inducing a different optimal partition. We enumerate these assignments while treating the intermediate devices as interchangeable.

%Unlike a centralized deployment, where devices can often be treated symmetrically, positions within a pipeline ring have distinct responsibilities. The head device hosts the embedding table, while the tail device hosts the final normalization and output projection layers; only the intermediate positions are interchangeable. This asymmetry affects both of the resources that govern partitioning. A device assigned to the head or tail must store additional model parameters and execute additional computations at every inference step. Consequently, both its fixed memory overhead, $c_i(r)$, and its fixed stage cost, $\alpha_i(r)$, depend on its role $r \in \{\text{head}, \text{mid}, \text{tail}\}$. 

%The challenge is therefore not merely deciding how many layers each device should receive, but also determining which devices should occupy the head and tail positions. For a replica containing $K$ devices, there are $K(K-1)$ possible head-tail assignments. Each assignment induces a different set of effective memory budgets and stage costs, and therefore a different optimal layer partition. Rather than enumerating all $K!$ device orderings, we exploit the symmetry of the intermediate positions and enumerate only the $K(K-1)$ admissible head-tail assignments, treating all remaining devices as interchangeable.

\paragraph{Memory limit} The first constraint on replica throughput is memory. Every device must store both the weights of its assigned layers and the key--value cache associated with all active requests. If a replica serves $s$ concurrent requests, device $i$ in role $r$ must satisfy
%??ref?? 
\vspace{-0.1cm}
\begin{equation} 
\vspace{-0.1cm}
    \ell_i w + s\,\ell_i \kappa + c_i(r) \le M_i, 
    \label{eq:memory} 
\end{equation} 
where $w$ denotes the per-layer weight footprint, $\kappa$ the per-layer key--value footprint
%??the same for all layers?? 
per request, and $M_i$ the usable memory capacity of the device. 
Eq.~\eqref{eq:memory} reveals an immediate tradeoff. Assigning more layers to a device increases both its weight footprint and its share of the key--value cache, reducing the number of concurrent requests that the device can support. Rearranging the constraint gives 
\[ s \le \frac{(M_i-c_i(r))/\ell_i - w}{\kappa}, \]
which is the maximum requests device $i$ can accommodate. Because this quantity decreases as $\ell_i$ grows, the replica's capacity is ultimately determined by the most memory-constrained device.
%??. 
Any imbalance in layer assignment therefore leaves memory unused on some devices while constraining concurrency on others. %??. 

The optimal partition avoids this imbalance by equalizing the memory-limited slot count ($s_{mem}$) across devices. This yields a water-filling solution in which $\ell_i \propto M_i-c_i(r)$, 
%\[ \ell_i \propto M_i-c_i(r) \] 
so devices with larger effective memory budgets receive proportionally more layers. Combining this relationship with the constraint $\sum_i \ell_i=L$ gives%??
\begin{equation} 
    s_{\mathrm{mem}}(S) = \frac{\widetilde{M}(S)-wL}{\kappa L}, 
    ~~~~
    \widetilde{M}(S) = \sum_{i\in S}\bigl(M_i-c_i(r_i)\bigr) 
    \label{eq:waterfillmem} 
\end{equation}
where $\widetilde{M}(S)$ denotes the aggregate memory available to the replica after accounting for role-specific overheads. 
Eq.~\ref{eq:waterfillmem} characterizes a \emph{memory-capacity bound} rather than the partition ultimately selected by the planner. It answers a simpler question: how many requests could the replica support if memory were the only limiting resource? Section~\ref{subsec:method:dominance} uses this bound to compare replicas and identify candidates that can be safely discarded.% before more expensive throughput evaluation is performed.

\paragraph{Concurrency limit} Memory determines how many requests a replica \emph{could} accommodate, but the maximum concurrency is ultimately set by throughput. As concurrency increases, the throughput available to each request falls because the available resources are shared among more in-flight requests. We therefore admit requests only while per-request throughput remains above the deployment target $\tau$ (i.e., minimum tokens per second). 
Let $\rho(S,s)$ denote the per-request throughput achieved by candidate $S$ when serving $s$ requests. Hence the largest admissible concurrency is
\begin{equation} 
    s_{con} = \max\{s : \rho(S,s) \ge \tau \}, 
    %~ 
    %\gamma(S) = \rho\!\bigl(S,s_{con}\bigr)\,s_{con} 
    \label{eq:concurrency} 
\end{equation} 
%where $s_{con}$ is the largest admissible concurrency level. % and $\gamma(S)$ is the aggregate throughput sustained at that concurrency limit. 
%Together, Eqs.~\ref{eq:waterfill} and~\ref{eq:replicavalue} define two complementary limits on a replica. 

Finally, the achievable serving request is therefore governed by whichever constraint becomes active first. A replica may exhaust its memory before throughput falls below $\tau$, or reach the throughput limit while memory remains available. In either case, the active constraint determines the concurrency level used when evaluating the replica. The overall throughput of a candidate $S$ is
\begin{equation} 
    \gamma(S) = \rho\!\bigl(S,s\bigr)\,s ,
    \quad
    s = \min\{s_{mem}, s_{con}\}
    \label{eq:replicathrput} 
\end{equation} 
%The memory bound determines how many requests the replica can hold, whereas the throughput target determines how many requests it can serve while meeting the required quality of service. The achievable serving request is therefore governed by whichever constraint becomes active first. A replica may exhaust its memory before throughput falls below $\tau$, or reach the throughput limit while memory remains available. In either case, the active constraint determines the concurrency level used when evaluating the replica.

\paragraph{The partition} Existing planners \cite{hu2022pipeedge,zhang2024edgeshard} obtain a layer assignment by DP over an ordered device sequence. We observe that the assignment admits a much simpler structure. Throughput is determined by the slowest pipeline stage, so any assignment that leaves one device slower than the others can be improved by shifting layers away from that device. At the optimum, therefore, all stages progress at the same rate. 

We model the execution time of device $i$ as an affine function of the number of layers assigned to it \vspace{-0.1cm}
\begin{equation} 
    t_i(\ell) = \alpha_i(r_i) + \beta_i \ell_i, 
    \label{eq:exetime}
\end{equation}
where $\alpha_i(r)$ captures the fixed per-token latency cost associated with role $r_i$ and $\beta_i$ is the measured per-layer latency cost of the device. Because $t_i$ increases monotonically with $\ell_i$, maximizing throughput is equivalent to minimizing the slowest stage, which occurs when all devices attain a common stage time $t^\star$. The layer assignments therefore satisfy $\ell_i =(t^\star-\alpha_i(r))/(\beta_i).$
%\[ \ell_i =(t^\star-\alpha_i(r))/(\beta_i). \]
Combining this condition with the conservation constraint $\sum_i \ell_i=L$ yields
\begin{equation} 
    L=\sum_i \big(\frac{t^\star-\alpha_i(r_i)}{\beta_i}\big) 
    \Rightarrow 
    t^\star = \frac{ L + \sum_i \alpha_i(r_i)/\beta_i }{ \sum_i 1/\beta_i }
    %\ell_i = \frac{t^\star-\alpha_i(r_i)}{\beta_i}. 
    \label{eq:waterfilltime} 
\end{equation} 
The partition therefore follows in closed form. Once $t^\star$ is known, the layer counts are computed directly, after which they are rounded and repaired to satisfy both $\ell_{\min}$ and the memory constraint of Eq.~\ref{eq:memory}. Because the head and tail devices incur different fixed costs, both $\alpha_i(r)$ and $c_i(r)$ depend on the selected role assignment. Eq.~\ref{eq:waterfilltime} is therefore solved once for each of the $K(K-1)$ admissible head--tail assignments, and the assignment achieving the highest throughput is retained. %Finally, we use $\gamma$ as the overall throughput of a candidate.

\paragraph{Complexity}
Water filling evaluates Eq.~\ref{eq:waterfilltime} in $O(K)$ time for a replica of $K$ devices. DP instead requires $O(KL^2)$ time by considering every layer boundary at every device. Enumerating the $K(K-1)$ head--tail assignments yields $O(K^3)$ under water filling versus $O(K^3L^2)$ under DP, an $L^2$ advantage that is significant in practice because $L \gg K$.

%The practical benefit of the water-filling formulation is that partitioning no longer depends on the number of layers. Evaluating Eq.~\ref{eq:waterfilltime} requires only two linear passes over the devices: one to accumulate $\sum_i \alpha_i/\beta_i$ and $\sum_i 1/\beta_i$, and one to emit the layer assignments, followed by a repair pass to enforce the constraints. The resulting cost is therefore $O(K)$ for a replica of $K$ devices.

%DP approaches scale differently. They maintain a state for each of the $L$ layer-boundary positions at every device and evaluate the admissible predecessors of each state, yielding $O(KL^2)$ complexity. Because head--tail assignments must still be enumerated, both methods are multiplied by the same $K(K-1)$ factor. Partitioning a replica therefore costs $O(K^3)$ under water filling, compared with $O(K^3L^2)$ under dynamic programming. The improvement is proportional to $L^2$, a substantial reduction in practice because modern LLMs contain far more layers than a deployment contains devices.

%This asymptotic advantage translates directly into planning time. Replacing dynamic programming with water filling reduces end-to-end planning time by $15.1\times$ on the 8-device deployment and by $29.7\times$ on the 30-device deployment. The observed speedup is smaller than the theoretical ratio because layer partitioning is only one component of the overall planning pipeline.

\subsection{A type-count representation for replicas}
\label{subsec:method:columns}

The observation that makes the outer search space manageable is that devices of the same hardware type are interchangeable. Devices sharing a type have the same memory budget and the same measured performance profile, so any two replicas that draw the same number of devices from each type receive identical throughput estimates and therefore identical scores. The planner need not distinguish individual devices. 

Since a replica is fully characterized by the devices assigned to it, we represent a candidate device allocation for a replica using a type-count vector $\mathbf{c}=(c_1,\dots,c_t)$
where $c_t$ denotes the number of devices drawn from type $t$ and $0\le c_t\le m_t$, with $m_t$ the number of available devices of that type. The search therefore moves from individual device subsets to configurations of device types. This representation reduces the number of candidates from the $2^N-1$ non-empty subsets of an $N$-device fleet to
\begin{equation} 
    \prod_{t=1}^{T}(m_t+1)-1, 
    \label{eq:columnspace} 
\end{equation}
which counts the non-empty type-count vectors. Candidates containing more than $K_{\max}$ devices are infeasible by construction and are discarded without evaluation.

\subsection{Search space pruning}
\label{subsec:method:dominance}

Even after the reduction of Eq.~\ref{eq:columnspace}, the search space remains large. We therefore identify \emph{dominated} candidates. A candidate is dominated if it can be decomposed into valid sub-candidates whose combined throughput is no smaller than its own. Such a candidate can be discarded.  %because whatever throughput it achieves can already be obtained by its decomposition. 
Dominance is evaluated recursively from previously generated candidates, avoiding explicit enumeration of all decompositions.

\paragraph{Discarding a dominated candidate is lossless.} Replacing a dominated candidate with the feasible replicas of its best decomposition preserves feasibility because both consume the same fleet resources. Since the decomposition provides at least as much aggregate throughput in the candidate's role, and the objective depends only on those aggregates, the replacement cannot increase congestion. An optimal placement therefore exists that excludes the dominated candidate, allowing it to be removed without loss of optimality. This argument does not extend to per-request metrics such as TPOT.%, where replacing one replica with several, changes the latency experienced by individual requests.

\begin{comment}
\paragraph{Discarding a dominated candidate is lossless.}
Suppose an optimal placement $X^\star$ contains candidate $\mathbf{a}$ in role $r$. Let $X'$ replace that replica with the feasible parts $\mathbf{b},\mathbf{c},\dots$ attaining the maximum in Eq.~\ref{eq:dominance}. The replacement preserves feasibility and cannot worsen the objective. Feasibility follows immediately because $\mathbf{b}+\mathbf{c}+\cdots=\mathbf{a}$: the replacement consumes exactly the same number of devices of each type and therefore remains a valid partition of the fleet. Its throughput is no smaller either, since Eq.~\ref{eq:dominance} guarantees 
\[ v_r(\mathbf{b})+v_r(\mathbf{c})+\cdots-v_r(\mathbf{a}) \ge 0. \] 
All other replicas remain unchanged. Because the objective $W$ depends on a placement only through $PS$ and $DS$ and is non-increasing in both, increasing throughput cannot increase congestion: a larger $PS$ can only reduce $C_P$ and increase $T_D$, while a larger $DS$ can only reduce $C_D$. Therefore, 
\[ W(X') \le W(X^\star). \] 
An optimal placement thus exists that does not use $\mathbf{a}$, and $\mathbf{a}$ may be discarded without loss of optimality. This argument relies on the objective depending only on aggregate Prefill and Decoder throughput. Per-request metrics such as TPOT do not possess this property: replacing one replica with several changes the latency experienced by individual requests, and the dominance argument no longer applies.
\end{comment}

\paragraph{A dominated candidate is limited by latency, not by memory}
Dominance alone saves search rather than evaluation, since a candidate's throughput must already be known before domination can be tested. The key observation is a memory ledger. Every replica stores a complete copy of the model weights, so decomposing a device set $\mathbf{a}$ into $\mathbf{b}+\mathbf{c}$ pays the weight cost twice. From Eq.~\ref{eq:waterfillmem}, \vspace{-0.1cm}
\begin{equation} 
\vspace{-0.1cm}
    s_{\text{mem}}(\mathbf{b})+s_{\text{mem}}(\mathbf{c}) = s_{\text{mem}}(\mathbf{a})-w/\kappa. 
    \label{eq:ledger} 
\end{equation} 
A decomposition therefore begins with fewer slots than the candidate it replaces. If it nevertheless achieves higher throughput, the gain must come from a higher service rate rather than additional memory. Dominance thus identifies candidates whose throughput is limited by latency, not memory.

\paragraph{Extending the rule to larger candidates.}
Discarding a dominated candidate saves search but not evaluation, since its value must already be known to test the domination. The real benefit arises if dominance also propagates to larger candidates. Let a set of device $d$ be added to $\mathbf{a}$, and suppose the same device is instead placed within one part of its dominating decomposition. 
\begin{comment}
Writing $\Delta_d^v(X)=v(X{+}d)-v(X)$, 
\begin{equation} 
    v(\mathbf{b}) + v(\mathbf{c}{+}d) = \underbrace{v(\mathbf{b})+v(\mathbf{c})}_{\ge\,v(\mathbf{a})} + \Delta_d^v(\mathbf{c}), 
    ~ 
    v(\mathbf{a}{+}d) = v(\mathbf{a}) + \Delta_d^v(\mathbf{a}), 
    \label{eq:extend} 
\end{equation} 
so the decomposition continues to dominate whenever $\Delta_d^v(\mathbf{c}) \ge \Delta_d^v(\mathbf{a})$. Intuitively, this is exactly the regime identified by dominance. 
\end{comment}
A dominated candidate has already grown beyond the point where additional memory translates into useful concurrency. 
%A dominated candidate is limited by latency, not by memory. 
Adding another device therefore contributes mainly an extra stage, whereas a smaller replica can still convert the added memory into throughput. The additional device is consequently worth more in the decomposition than in the already dominated candidate. 

A general proof is significantly complicated by role-dependent costs, heterogeneous devices, and the fact that a decomposition and its parent candidate need not operate in the same regime. We therefore adopt this extension as an assumption used during candidate generation rather than as a theorem.
While violating the assumption may invalidate the optimality guarantee, our experimental results suggest that the resulting planner still produces high-quality solutions while significantly reducing solving time compared to existing methods.
%The assumption does not affect the exactness of the outer solver: pruning a dominated candidate is lossless, and exhaustive enumeration remains available whenever desired. 
The assumption affects only which candidates are generated. The outer solver remains exact over those candidates, and removing a dominated candidate is itself lossless (see Section \ref{subsec:method:dominance}.a).
The extension simply avoids generating descendants of candidates that have already been shown uncompetitive. In Section~\ref{sec.eval}, we show that exhaustive enumeration and dominance pruning produce identical survivor sets, Pareto frontiers, and final placements whenever exhaustive comparison is affordable.

\subsection{Placement method} 
\label{subsec:method:outer} 
The surviving candidates form a \textbf{bounded multidimensional knapsack}: choose how many replicas of each surviving type-count vector to instantiate, and in which role, subject to the available devices of each type. Since the objective $W$ depends on a placement only through $(PS,DS)$, the problem reduces to enumerating the Pareto frontier of achievable throughput pairs and evaluating $W$ on that frontier. %Eq.~\ref{eq:objective} on that frontier. 

We consider two exact formulations. The first is a multi-dimension knapsack DP \cite{kellerer2004multidimensional} whose states are device-consumption vectors $\mathbf{u}\le(m_1,\dots,m_T)$. Each transition adds one surviving candidate in one of the two roles and updates the reachable $(PS,DS)$ frontier. This DP is different from layer-partitioning DP. The second is a mixed-integer program over replica counts $x^r_{\mathbf{c}}$: 
\begin{align} 
    \min\quad & W(PS, DS) \nonumber\\ \text{s.t.}\quad & \sum_{r}\sum_{\mathbf{c}} x^{r}_{\mathbf{c}}\,c_t \le m_t \quad \forall t, \label{eq:milp:cap}\\ & PS = \sum_{\mathbf{c}} x^{P}_{\mathbf{c}}\,\gamma_P(\mathbf{c}), \qquad DS = \sum_{\mathbf{c}} x^{D}_{\mathbf{c}}\,\gamma_D(\mathbf{c}). 
    \label{eq:milp:agg} 
\end{align} 

The two formulations trade time for memory. The DP is faster because it constructs the entire Pareto frontier in a single forward pass and remains exact by construction. However, its speed comes at the cost of working memory: the state table spans $\prod_t (m_t+1)$ entries and therefore grows exponentially with the number of device types. This is often acceptable for heterogeneous edge fleets, where the number of types is small even when the number of devices is large, but it eventually becomes the limiting resource.

The mixed-integer program makes the opposite tradeoff. By avoiding the state table altogether, its memory footprint remains modest as the type count grows. However, the cost is runtime: the Pareto frontier must be recovered through repeated solves, each corresponding to a different scalarization of $(PS,DS)$. We therefore use the DP whenever its state space fits comfortably in memory and fall back to the mixed-integer program once memory becomes the bottleneck. Section~\ref{subsec:eval:planner} reports the memory crossover on our deployments.

\section{Evaluation and Discussion}\label{sec.eval}
% NOTE: numbers marked \pending{} are not yet measured. Do not quote them.
\providecommand{\pending}[1]{\textbf{[pending: #1]}}

Our evaluation proceeds in four steps. We first investigate which existing solutions can serve as meaningful baselines in our deployment regime and explain why others cannot be compared directly (Section~\ref{subsec:eval:baselines}). We then validate the per-replica cost model on which dominance pruning depends (Section~\ref{subsec:eval:accuracy}). Next, we compare \textsc{Spade} with placement planners of the selected baselines in both solution quality and planning time, isolating the contribution of the dominance rule (Section~\ref{subsec:eval:planner}). Finally, we study how the queueing-aware objective affects the resulting deployment as the model, workload, arrival rate, and fleet size vary (Section~\ref{subsec:eval:objective}).

\subsection{Experiment setup}
\label{subsec:eval:setup}

\begin{table}[t]
    \centering
    \caption{Device information. (U.M: Unified Memory)}
    \label{tab:dev}
    \begin{tabular}{|c|c|c|c|c|c|}
        \hline
         Device Name & VRAM & RAM & U.M& Type.ID & N.Dev\\\hline
         RTX5070 & 12GB & 64GB & No & Type.1 & 1\\
         Ryzen 7 9700X & & & & &\\\hline
         RTX3060M & 6GB & 64GB & No & Type.2 & 1\\
         Ryzen 5 5800H & & & & &\\\hline
         Apple M2 Max & 22GB & 32GB & Yes & Type.3 & 2\\\hline
         Apple M1 & 12GB & 16GB & Yes & Type.4 & 2\\\hline
         DGX Spark* & 12GB & 128GB & Yes & Type.5 & 2\\\hline
    \end{tabular}
\end{table}

Our testbed consists of eight devices spanning five hardware types and connected through a local-area network (Table~\ref{tab:dev}). To emulate a resource-constrained edge device, we limit the DGX Spark to 12\,GB of VRAM. %Following E2LLM \cite{le2026e2llm}, we use network bandwidth that is sufficient for communication overhead to remain secondary to pipeline execution time (about 900Mbps). 
Following E2LLM \cite{le2026e2llm}, we deploy our testbed on a Local Area Network (about 900Mbps) to reduce the affect of transmitting time on pipeline paralellism.
We evaluate two contrasting models: DeepSeek\_R1\_Distill\_Qwen\_14B\_Q6\_K and Qwen3-30B-A3B-Q4\_K\_M, referred to as \textit{DeepSeekR1} and \textit{Qwen3MoE}. DeepSeekR1 stores 11.3\,GB of weights but activates all 14B parameters through dense FFNs \cite{deepseekai2025deepseekr1incentivizingreasoningcapability}, whereas Qwen3MoE stores 17.3\,GB of weights but activates only about 3B parameters through a Mixture-of-Experts architecture \cite{qwen3technicalreport}. The two models therefore represent latency- and memory-constrained deployment regimes, respectively. Workloads are drawn from the LZByteDance \cite{ling2025longreason} and IELTS \cite{chillies_dpo_ielts_writing} datasets which together represent different $N_P$/$N_D$ ratios as in Table \ref{tab:RatioPrefillDecoderDataset}.
Moreover, the service target is set to $\tau=15$ tokens/s per request, doubling the average human reading speed \cite{brysbaert2019many}. Request inter-arrival times follow a normal distribution with mean $T_a$ and standard deviation $\sigma = 0.1T_a$, truncated at zero. %Table~\ref{tab:RatioPrefillDecoderDataset} gives the %average input and output lengths of each workload.

\begin{table}[]
    \centering
    \caption{Average number of input and generated tokens.}
    \vspace{-0.2cm}
    \label{tab:RatioPrefillDecoderDataset}
    \begin{tabular}{|c|c|c|c|c|}
        \hline
         Model & Dataset &$N_P$ & $N_D$ & ratio\\\hline
         DeepseekR1 & LZByteDance  &576 & 579 & 0.99\\\cline{2-5}
         &IELTS & 2282 & 1173 & 1.94\\\hline
         Qwen3Moe & LZByteDance  &576 & 746 & 0.77\\\cline{2-5}
         Q4\_K\_M&IELTS & 2291 & 1957 & 1.17\\\hline
    \end{tabular}
    \vspace{-0.3cm}
\end{table}

\subsection{Baselines chosen}
\label{subsec:eval:baselines}

\textbf{Systems that cannot execute in this regime.}
Several published systems cannot be evaluated directly on our testbed. PipeEdge targets encoder models without a key--value cache, and its partitioning cost model does not extend to autoregressive decoding. In addition, HexGen, HexGen-2, HexGen-3, ThunderServe and Splitwise depend on CUDA and NCCL, which are unavailable on Mac devices and edge fleet on LAN. We therefore cannot compare against their full implementations. Instead, we compare with their planners and objectives, as mentioned below. %search strategies and objectives rather than their implementations. 

\textbf{Objectives evaluated inside our planner.}
Placement objectives and placement searches solve different problems, and evaluating them together confounds the effect of one with the other. Therefore, within our own planner, we evaluate the following \emph{placement objectives}: finishing time (from HexGen \cite{jiang2024hexgen} - FT), role-pinned disaggregation (from Splitwise \cite{patel2024splitwise}, called adapted Splitwise - SPW), and the congestion objective in E2LLM. %--- within our own planner while holding the search procedure fixed. 
For SPW, we derive a role-pinned policy from the deployment choices reported in Splitwise, where the higher-throughput resources are consistently assigned to Prefill and lower-throughput to Decode. Differences in the resulting placements can then be attributed to the objective alone rather than to differences in the implementation. These experiments should therefore be understood as objective-level ablations.
%rather than implementation ports.

\textbf{Search methods compared directly.}
To isolate the effect of the search algorithm, we compare three alternatives: a genetic algorithm (GA) combined with our water-filling partitioner, a GA paired with the layer-partitioning DP used in E2LLM \cite{le2026e2llm}, and the tabu search employed by ThunderServe \cite{jiang2025thunderserve}. All methods are run with the parameters reported by their authors; no search-specific tuning was performed for our fleet.

Moreover, prima.cpp \cite{li2026prima} is the closest published system that executes on our hardware.
Halda as its planner relies on pipeline-ring parallelism (PRP), in which weight transfers overlap computation through multiple execution rounds ($k>1$). Rebuilding prima.cpp with a GPU backend and reproducing Halda's parameter selection on our devices showed that this mechanism never activates in our setting. For PRP to hide storage latency, the device compute rate must satisfy
\vspace{-0.1cm}
\begin{equation}
\vspace{-0.1cm}
\textit{rate} \le (N-1)\,I\,\textit{disk\_bw}
\end{equation}
where $I$ is the model arithmetic intensity and $\textit{disk\_bw}$ the storage bandwidth. 
%Importantly, $k$ cancels from the derivation, so increasing the number of rounds cannot move a deployment into the PRP regime. 
For the Llama-3-70B\_Q4\_K\_M model used in their original evaluation, the bound is approximately 60\,GFLOPS per device on a four-device deployment ($I=3.32$, SSD $\textit{disk\_bw}=6GB/s$), whereas the slowest CPU in our fleet reaches 66.4\,GFLOPS and the fastest GPU 87{,}207\,GFLOPS, measured by their tool. The overlap condition is therefore violated throughout our fleet, causing prima.cpp to fall back to a single-round pipeline with only one microbatch in flight. While still pipeline-parallel, this execution regime no longer benefits from pipeline filling or weight-fetch overlap, both of which are fundamental to efficient pipeline parallelism \cite{zhang2024edgeshard,hu2022pipeedge,le2024optimal}. Since prima.cpp operates in a substantially inefficient regime from the studies in this work, we do not include it in the end-to-end comparison.

\subsection{Pipeline parallelism performance prediction}
\label{subsec:eval:accuracy}

The dominance rule prunes candidates without evaluating them, and the outer solver ranks placements that it never needs to execute. Both therefore depend on the accuracy of the per-replica cost model. Following E2LLM \cite{le2026e2llm} and EdgeShard \cite{zhang2024edgeshard}, we model a fully populated pipeline by its bottleneck stage: \vspace{-0.1cm}
\begin{equation} 
    \vspace{-0.1cm}
    TPOT = T_{\mathrm{slowest}} \times N_{\mathrm{stage}}. 
    \label{eq:ppperf} 
\end{equation} 

\begin{figure}[b]
    \centering
    \vspace{-0.2cm}
    \includegraphics[width=0.6\linewidth]{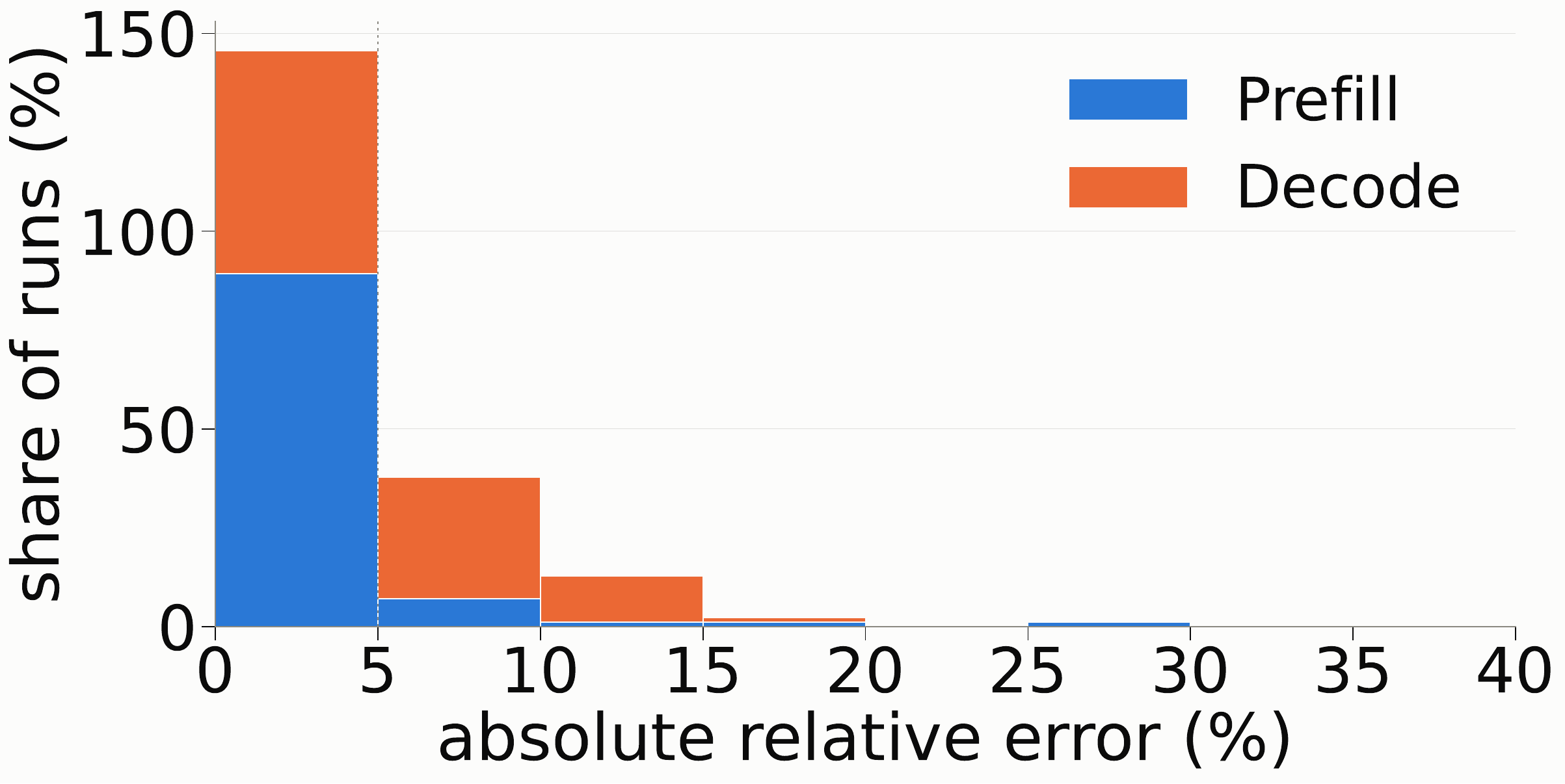}
    \vspace{-0.1cm}
    \caption{Error distribution of pipeline parallelism performance prediction. Each stage's distribution sums to 100\%.}
    \label{fig:Pipelinepred}
    %\vspace{-0.1cm}
\end{figure}

To validate Eq.~\ref{eq:ppperf}, we sampled deployments from our fleet, including non-optimal placements.%, measured their stage times, and compared the predicted throughput against measured throughput. 
When evaluated on measured stage times and performance, Eq.~\ref{eq:ppperf} predicts throughput with less than 0.5\% error across all deployments, confirming that throughput is well explained by the bottleneck-stage model. 
%We next validated the stage-time predictor itself. Comparing Eq.~\ref{eq:exetime} against measured execution times, the prediction error on the bottleneck stage was typically below 5\%, while non-bottleneck stages occasionally reached errors of up to 50\%. Despite these discrepancies, the predictor identified the correct bottleneck stage in every experiment. 
%The larger errors on non-bottleneck stages arise from device-level power management. During profiling, workloads execute continuously, whereas in a running pipeline faster stages often idle while waiting for the bottleneck. On devices such as the RTX3060M and Apple M2 Max, these idle periods trigger different power states and therefore different execution rates. Because throughput is governed by the bottleneck stage, these variations have little impact on overall prediction accuracy. Consequently, using predicted rather than measured stage times yields throughput estimates with errors comparable to those of the bottleneck-stage prediction itself (Figure~\ref{fig:Pipelinepred}).
We next combine the stage-time predictor (Eq.~\ref{eq:exetime}) with the bottleneck-stage model of Eq.~\ref{eq:ppperf} to estimate replica throughput. The predictions achieve an average error of 5.22\% across all deployments. Figure~\ref{fig:Pipelinepred} shows the error distribution.

\subsection{Planner comparison}
\label{subsec:eval:planner}

Table~\ref{tab:prunecomp} evaluates the effect of dominance pruning on synthetic 30 devices derived from our 8-device testbed by replicating the observed hardware types with varying memory capacities and compute throughput. The evaluations are based on the water-filling method. Without pruning, every candidate must be generated and priced before the outer solver runs. With pruning, only candidates that survive the dominance test are evaluated. The savings are substantial. At $T=5$ device types, exhaustive generation evaluates 16554 candidates and requires 153\,s for the whole process, whereas the pruned generator evaluates only 238 candidates and completes everything in 0.44\,s, a $347\times$ reduction in runtime. At $T=10$, the gap widens from 35\,252\,s (more than nine hours) to 11.24\,s, a $3\,136\times$ reduction, with only 3\,710 of 1\,045\,682 candidates ever evaluated. The advantage grows with the number of device types, which is precisely the dimension along which the search space grows exponentially. 
These savings come with no observable loss in solution quality because at both scales, exhaustive and pruned generation produce identical survivor sets, identical Pareto frontiers, and the same final placement. This is the empirical validation referred to in Section~\ref{subsec:method:dominance}: although the extension rule is not proven lossless in general, exhaustive enumeration confirms that it preserves all candidates required to recover the optimum.% wherever such verification remains computationally affordable.

\begin{table}[b]
    \centering
    \setlength{\tabcolsep}{4pt}
    \vspace{-0.1cm}
    \caption{Performance comparison between exhaustive and pruned generations ($N=30$, planner DP). Total time also includes time for other operations, e.g, file writing, comparing}
    \vspace{-0.2cm}
    \label{tab:prunecomp}
    \begin{tabular}{|l|l|rrrrr|}
        \hline
        T & Generator & candidates & survivors & generation & Planner & total\\\hline
        \multirow{2}{*}{$5$}
          & exhaustive & 16\,554    & 79  & 152.0\,s   & 389.3\,ms & 152.9\,s\\
          & pruned     & \textbf{238}  & 79  & 338.3\,ms  & 3.2\,ms   & \textbf{0.44\,s}\\\hline
        \multirow{2}{*}{$10$}
          & exhaustive & 1\,045\,682 & 750 & 25\,949\,s & 8\,896\,s & 35\,252\,s\\
          & pruned     & \textbf{3\,710} & 750 & 8.8\,s     & 2.2\,s    & \textbf{11.2\,s}\\\hline
    \end{tabular}
    \vspace{-0.2cm}
\end{table}

\begin{table}[t]
    \centering
    \setlength{\tabcolsep}{4pt}
    \caption{Performance of planners. Lower is better}
    \vspace{-0.2cm}
    \label{tab:plannersearchcomp}
    \begin{tabular}{|l|cc|cc|}
        \hline
        \multirow{2}{*}{Search} & \multicolumn{2}{c|}{N=8, T=5} & \multicolumn{2}{c|}{N=30, T=5}\\
        \cline{2-5}
         & time & gap (\%) & time & gap (\%)\\\hline
        \textsc{Spade} (knapsack DP) &  \textbf{109.4\,ms} & 
        \textbf{best} & \textbf{441.3\,ms} & \textbf{best}\\
        \textsc{Spade} (MILP, full Pareto points) & 223.0\,ms& \textbf{best} & 11.3s & \textbf{best}\\\hline
        %MILP, 16-point grid          &  292.2\,ms &        $+6.2$ & 680.0\,ms &        $+3.9$\\
        %MILP, 64-point grid          &  594.2\,ms & \textbf{best} &   2.28\,s & \textbf{best}\\
        %MILP, 256-point grid         &    1.72\,s & \textbf{best} &   8.10\,s & \textbf{best}\\\hline
        tabu search (ThunderServe)          &  384.4\,ms & \textbf{best} &   13.4\,s &       $+20.4$\\
        GA + water filling           &  548.0\,ms & \textbf{best} &    231\,s &       $+18.8$\\
        GA + replica DP (E2LLM)      &    8.26\,s & \textbf{best} &   6855\,s &       $+19.7$\\\hline
    \end{tabular}
    \vspace{-0.2cm}
\end{table}

\paragraph{Search methods} Table \ref{tab:plannersearchcomp} compares the placement searches on two rosters: the physical fleet and a 30-device roster constructed by replicating the measured device types. On the physical fleet, all methods reach the same optimum and differ only in planning cost. Our planner completes in 109.4\,ms, compared with 384.4\,ms for tabu search, 548.0\,ms for the genetic algorithm with water-filling partitioning, and 8\,257\,ms for the genetic algorithm paired with the layer-partitioning DP. At eight devices the search space is sufficiently small that solution quality is no longer discriminative, and we report this rather than selecting only the larger instance. 

The picture changes at 30 devices. Our planner reaches the verified optimum in 441.3\,ms. The mixed-integer formulation also recovers the optimum when the weight grid is sufficiently fine, requiring 31\,ms per Pareto point $(PS,DS)$. Recovering the full frontier therefore takes 11.3\,s, roughly $25.6\times$ longer than the DP. This asymmetry is structural rather than an implementation artifact: the DP recovers the entire frontier in a single backward pass, whereas the MILP must be solved separately for each Pareto point. The heuristic methods separate both in runtime and solution quality. The GA yields 18.8\% higher congestion after 230.7\,s; replacing water filling with the layer-partitioning DP increases runtime to 6\,855\,s while still yielding 19.7\% higher congestion. Tabu search completes in 13.4\,s but yields 20.4\% higher congestion. Interestingly, varying the parameters of tabu search does not materially reduce the gap, and likewise for GA, a typical behavior of heuristic searches that provide no optimality guarantees.
%, where additional search effort does not necessarily translate into better solutions. 
Optimal planning is hence tractable at this scale: \textsc{Spade} reaches the optimum in milliseconds, while the heuristic ones are either slower, less accurate, or both.

%The 30-device roster is synthetic, so these results support claims about planning time and solution quality under the cost model rather than deployed throughput. We also report the mixed-integer program both with and without candidate generation, since it consumes the same candidate set produced by our generator.

\paragraph{Memory} \label{subsec:eval:planner:memory} The two exact formulations exhibit the tradeoff predicted in Section~\ref{subsec:method:outer}. The multidimensional-knapsack DP is the faster solver, but its dense state table grows exponentially with the number of device types. On our largest roster, it requires 4.75\,GB of memory. The mixed-integer program makes the opposite tradeoff, reproducing the same frontier using only 122\,MB while requiring longer solve times. With larger device types, for example 50 devices in 20 types, the DP no longer fits in memory, leaving the mixed-integer formulation as the only practical exact solver.

\subsection{Placement objective comparison}
\label{subsec:eval:objective}

%The preceding results isolate the effect of the search procedure by holding the objective fixed. 
We now hold the planner fixed and vary only the objective, so any differences in the resulting deployments are attributable to the objective itself. 
%All four objectives are optimized by the same solver over the same candidate set. 
Following E2LLM, we evaluate them across four dimensions: model architecture, arrival rate, workload, and fleet scale. 
%For the scalability study, the 4-device fleet consists of one Type.1, one Type.2, and two Type.4 devices; the 6-device fleet additionally includes one Type.3 and one Type.5 device; and the 8-device fleet uses all devices. 
%Although each objective selects its own placement, every placement is evaluated using the same congestion metric of Eq.~\ref{eq:objective}. We additionally report aggregate decode throughput because $W$ is clamped at zero once the offered load can be sustained, at which point throughput continues to distinguish the resulting deployments. 
%The finishing-time objective is reported as defined, including its negative values in under-loaded regimes. 
Table~\ref{tab:objective} summarizes results of the four objectives across all evaluation dimensions.
In Tables~\ref{tab:dep-model}, \ref{tab:dep-workload}, and \ref{tab:dep-arrival}, \textbf{R} denotes the replica role (P = Prefill, D = Decode), \textbf{Sl.} the total slot count, and \textbf{G-C} the numbers of GPU and CPU layers assigned to the replica. The first device is always the head and the last device is always the tail.
For role-pinned placement (SPW), we do not apply dominance pruning because the two techniques interact poorly. 
It should be noted that in Table \ref{tab:objective}, $PS$ and $DS$ are measured on average. Hence, even with the same deployment plan, the average performance may be different depending on the system saturation level.

%The preceding results isolate the effect of the search procedure by holding the objective fixed. We now reverse that perspective: the planner remains unchanged, and only the objective varies. Any differences in the resulting placements can therefore be attributed to the objective itself rather than to the search. To ensure a fair comparison, all four objectives are optimized by the same solver over the same candidate set. Following E2LLM, we evaluate the objectives along four dimensions: model architecture, arrival rate, workload, and fleet scale. For the scalability study, the 4-device fleet consists of one Type.1 device, one Type.2 device, and two Type.4 devices. The 6-device fleet additionally includes one Type.3 and one Type.5 device, while the 8-device fleet uses all physical devices in the testbed.

\begin{table*}[t]
    \centering
    %\footnotesize
    \setlength{\tabcolsep}{2.5pt}
    \caption{Measured metrics of the placement each objective selects. $PS$/$DS$ are per-request prefill/decode throughput (tokens/s); $W_{\max}$ is worst waiting time (s). Higher $PS$/$DS$ and lower $W_{max}$ are better; the best in each row is bold.}
    \vspace{-0.2cm}
    \label{tab:objective}
    \begin{tabular}{|l|ccc|ccc|ccc|ccc|}
        \hline
        & \multicolumn{3}{c|}{\textsc{Spade}} & \multicolumn{3}{c|}{FT} & \multicolumn{3}{c|}{E2LLM} & \multicolumn{3}{c|}{SPW}\\
        Configuration & $PS$ & $DS$ & $W_{\max}$ & $PS$ & $DS$ & $W_{\max}$ & $PS$ & $DS$ & $W_{\max}$ & $PS$ & $DS$ & $W_{\max}$ \\\hline
        \multicolumn{13}{|l|}{\emph{Model} --- 8 dev., LZByteDance, $100$~ms}\\\hline
        Qwen3-MoE    & 183.2 & 28.9 & \textbf{148} & 366.3 & 31.3 & 155 & 366.3 & 31.3 & 155 & 1315.1 & 25.4 & 169\\
        %\quad\textit{\scriptsize err.\ \%} & \scriptsize +38 & -18 & --- & \scriptsize +51 & -13 & --- & \scriptsize +51 & -13 & --- & \scriptsize +22 & -21 & ---\\
        DeepSeekR1   & 198.6 & 16.2 & \textbf{324} & 761.1 & 15.2 & 791 & 761.1 & 15.2 & 791 & 752.9 & 19.5 & 1014\\
        %\quad\textit{\scriptsize err.\ \%} & \scriptsize +9.1 & +2.9 & --- & \scriptsize +6.9 & +13 & --- & \scriptsize +6.9 & +13 & --- & \scriptsize +23 & +35 & ---\\
        \hline
        \multicolumn{13}{|l|}{\emph{Workload} --- 8 dev., Qwen3-MoE, $100$~ms}\\\hline
        IELTS        & 694.8 & 23.4 & \textbf{372} & 394.7 & 25.6 & 400 & 394.7 & 25.6 & 400 & 1510.9 & 21.5 & 458\\
        %\quad\textit{\scriptsize err.\ \%} & \scriptsize -4.3 & -6.0 & --- & \scriptsize +25 & -5.8 & --- & \scriptsize +25 & -5.8 & --- & \scriptsize +4.1 & -17 & ---\\
        LZByteDance  & 183.2 & 28.9 & \textbf{148} & 366.3 & 31.3 & 155 & 366.3 & 31.3 & 155 & 1315.1 & 25.4 & 169\\
        %\quad\textit{\scriptsize err.\ \%} & \scriptsize +38 & -18 & --- & \scriptsize +51 & -13 & --- & \scriptsize +51 & -13 & --- & \scriptsize +22 & -21 & ---\\
        \hline
        \multicolumn{13}{|l|}{\emph{Arrival interval} --- 8 dev., Qwen3-MoE, LZByteDance}\\\hline
        $100$~ms     & 183.2 & 28.9 & \textbf{148} & 366.3 & 31.3 & 155 & 366.3 & 31.3 & 155 & 1315.1 & 25.4 & 169\\
        %\quad\textit{\scriptsize err.\ \%} & \scriptsize +38 & -18 & --- & \scriptsize +51 & -13 & --- & \scriptsize +51 & -13 & --- & \scriptsize +22 & -21 & ---\\
        $1000$~ms    & 183.8 & 28.9 & \textbf{59} & 518.7 & 31.2 & 72 & 518.7 & 31.2 & 72 & 1210.0 & 25.4 & 69\\
        %\quad\textit{\scriptsize err.\ \%} & \scriptsize +38 & -17 & --- & \scriptsize +6.9 & -13 & --- & \scriptsize +6.9 & -13 & --- & \scriptsize +33 & -17 & ---\\
        $3000$~ms    & 213.2 & 59.0 & 7 & 508.7 & 46.8 & \textbf{1} & 213.2 & 59.0 & 7 & 1210.1 & 33.1 & 2\\
        %\quad\textit{\scriptsize err.\ \%} & \scriptsize +31 & -59 & --- & \scriptsize +9.0 & -42 & --- & \scriptsize +31 & -59 & --- & \scriptsize +33 & -40 & ---\\
        \hline
        \multicolumn{13}{|l|}{\emph{Cluster size} --- Qwen3-MoE, IELTS, $100$~ms}\\\hline
        4 dev.       & 264.1 & 21.5 & \textbf{1477} & 235.1 & 18.7 & 1729 & 235.1 & 18.7 & 1729 & 1475.6 & 25.0 & 3838\\
        %\quad\textit{\scriptsize err.\ \%} & \scriptsize +4.5 & -2.6 & --- & \scriptsize +1.9 & -5.9 & --- & \scriptsize +1.9 & -5.9 & --- & \scriptsize +6.6 & +35 & ---\\
        6 dev.       & 235.5 & 31.5 & \textbf{598} & 315.6 & 32.7 & 666 & 315.6 & 32.7 & 666 & 1476.6 & 17.2 & 804\\
        %\quad\textit{\scriptsize err.\ \%} & \scriptsize +1.8 & -18 & --- & \scriptsize -1.5 & -5.4 & --- & \scriptsize -1.5 & -5.4 & --- & \scriptsize +6.6 & -3.0 & ---\\
        8 dev.       & 694.8 & 23.4 & \textbf{372} & 394.7 & 25.6 & 400 & 394.7 & 25.6 & 400 & 1510.9 & 21.5 & 458\\
        %\quad\textit{\scriptsize err.\ \%} & \scriptsize -4.3 & -6.0 & --- & \scriptsize +25 & -5.8 & --- & \scriptsize +25 & -5.8 & --- & \scriptsize +4.1 & -17 & ---\\
        \hline
    \end{tabular}
    \vspace{-0.2cm}
\end{table*}

\paragraph{Model change} We first vary the model while holding the fleet (eight devices), workload (LZByteDance), and arrival period (100\,ms) fixed. Our congestion objective outperforms all baselines on both models, but the magnitude of the improvement depends strongly on the model architecture. On Qwen3MoE, the maximum waiting times produced by FT, E2LLM, and Splitwise are only 4.7\%, 4.7\%, and 14.2\% higher than \textsc{Spade}, respectively (Table \ref{tab:objective}). Although these differences appear modest, waiting time accumulates quadratically with the request count% according to Eq.~\ref{eq:objective}
, causing small per-request discrepancies to translate into substantially different backlog growth under sustained load. The gap becomes much larger on DeepSeekR1, where FT, E2LLM, and Splitwise incur maximum waiting times that are 2.44$\times$, 2.44$\times$, and 3.13$\times$ as long as \textsc{Spade}, respectively. The contrast reflects the different operating regimes of the two models. Qwen3MoE activates only a small fraction of its parameters per token, allowing it to sustain a relatively high service rate. As concurrency increases, the deployment continues admitting requests until memory capacity becomes the limiting factor, as shown in column DS of Table \ref{tab:objective}. DeepSeekR1, in contrast, activates all model parameters through dense feed-forward networks, resulting in a substantially lower service rate. Consequently, the throughput target is reached before memory is exhausted (Eq. \ref{eq:waterfillmem}, \ref{eq:concurrency} and \ref{eq:replicathrput}), making the deployment limited by latency, rather than memory. Objectives that ignore phase-specific congestion are therefore penalized more severely on DeepSeekR1.

\begin{table}[t]
    \vspace{-0.1cm}
    \centering
    \setlength{\tabcolsep}{2pt}
    %\caption{Deployment plans when varying the model (eight devices, LZByteDance, $100$~ms).}
    \caption{Deployment plans when varying the model}
    \vspace{-0.2cm}
    \label{tab:dep-model}
    \begin{subtable}[t]{.325\linewidth}
        \centering
        \caption{Qwen3MoE / \textsc{Spade}.}
        \begin{tabular}{|c|c|c|c|}
            \hline
            R & Type & Sl. & G--C\\\hline
            \multirow{2}{*}{P} & Type.2 & \multirow{2}{*}{2} & 8--20\\
              & Type.4 &  & 20--0\\\hline
            \multirow{3}{*}{D} & Type.5 & \multirow{3}{*}{15} & 16--0\\
              & Type.5 &  & 16--0\\
              & Type.1 &  & 15--1\\\hline
            \multirow{3}{*}{D} & Type.3 & \multirow{3}{*}{12} & 22--0\\
              & Type.4 &  & 7--0\\
              & Type.3 &  & 19--0\\\hline
        \end{tabular}
    \end{subtable}\hfill
    \begin{subtable}[t]{.325\linewidth}
        \centering
        \caption{Qwen3MoE / \\FT $\equiv$ E2LLM.}
        \begin{tabular}{|c|c|c|c|}
            \hline
            R & Type & Sl. & G--C\\\hline
            \multirow{2}{*}{P} & Type.4 & \multirow{2}{*}{2} & 9--0\\
              & Type.3 &  & 39--0\\\hline
            P & Type.3 & 1 & 48--0\\\hline
            \multirow{3}{*}{D} & Type.5 & \multirow{3}{*}{15} & 16--0\\
              & Type.5 &  & 16--0\\
              & Type.1 &  & 15--1\\\hline
            \multirow{2}{*}{D} & Type.4 & \multirow{2}{*}{4} & 19--0\\
              & Type.2 &  & 5--24\\\hline
        \end{tabular}
    \end{subtable}\hfill
    \begin{subtable}[t]{.325\linewidth}
        \centering
        \caption{Qwen3MoE / Splitwise.}
        \begin{tabular}{|c|c|c|c|}
            \hline
            R & Type & Sl. & G--C\\\hline
            P & Type.1 & 1 & 25--23\\\hline
            \multirow{6}{*}{D} & Type.2 & \multirow{6}{*}{18} & 3--3\\
              & Type.4 &  & 3--0\\
              & Type.4 &  & 3--0\\
              & Type.3 &  & 10--0\\
              & Type.5 &  & 13--0\\
              & Type.5 &  & 13--0\\\hline
            D & Type.3 & 2 & 48--0\\\hline
        \end{tabular}
    \end{subtable}

    \vspace{0.1cm}
    \begin{subtable}[t]{.325\linewidth}
        \centering
        \caption{DeepSeekR1 / \textsc{Spade}.}
        \begin{tabular}{|c|c|c|c|}
            \hline
            R & Type & Sl. & G--C\\\hline
            P & Type.3 & 1 & 48--0\\\hline
            \multirow{3}{*}{D} & Type.5 & \multirow{3}{*}{9} & 15--0\\
              & Type.5 &  & 15--0\\
              & Type.1 &  & 18--0\\\hline
            \multirow{4}{*}{D} & Type.2 & \multirow{4}{*}{2} & 15--0\\
              & Type.4 &  & 6--0\\
              & Type.4 &  & 6--0\\
              & Type.3 &  & 21--0\\\hline
        \end{tabular}
    \end{subtable}\hfill
    \begin{subtable}[t]{.325\linewidth}
        \centering
        \caption{DeepSeekR1 / \\FT $\equiv$ E2LLM.}
        \begin{tabular}{|c|c|c|c|}
            \hline
            R & Type & Sl. & G--C\\\hline
            \multirow{2}{*}{P} & Type.2 & \multirow{2}{*}{2} & 13--0\\
              & Type.5 &  & 35--0\\\hline
            \multirow{2}{*}{D} & Type.5 & \multirow{2}{*}{6} & 24--0\\
              & Type.1 &  & 24--0\\\hline
            \multirow{4}{*}{D} & Type.4 & \multirow{4}{*}{2} & 5--0\\
              & Type.3 &  & 20--0\\
              & Type.3 &  & 20--0\\
              & Type.4 &  & 3--0\\\hline
        \end{tabular}
    \end{subtable}\hfill
    \begin{subtable}[t]{.325\linewidth}
        \centering
        \caption{DeepSeekR1 / Splitwise.}
        \begin{tabular}{|c|c|c|c|}
            \hline
            R & Type & Sl. & G--C\\\hline
            \multirow{2}{*}{P} & Type.5 & \multirow{2}{*}{2} & 13--0\\
              & Type.5 &  & 35--0\\\hline
            D & Type.3 & 1 & 48--0\\\hline
            \multirow{4}{*}{D} & Type.4 & \multirow{4}{*}{2} & 2--0\\
              & Type.2 &  & 15--0\\
              & Type.4 &  & 2--0\\
              & Type.1 &  & 29--0\\\hline
            D & Type.3 & 1 & 48--0\\\hline
        \end{tabular}
    \end{subtable}
    \vspace{-0.45cm}
\end{table}

\begin{table}[t]
    \centering
    \vspace{-0.1cm}
    \setlength{\tabcolsep}{2pt}
    %\caption{Deployment plans when varying the workload (eight devices, Qwen3-MoE, $100$~ms).}
    \caption{Deployment plans when varying the workload}
    \vspace{-0.2cm}
    \label{tab:dep-workload}
    \begin{subtable}[t]{.325\linewidth}
        \centering
        \caption{IELTS / \textsc{Spade}.}
        \begin{tabular}{|c|c|c|c|}
            \hline
            R & Type & Sl. & G--C\\\hline
            P & Type.3 & 1 & 48--0\\\hline
            \multirow{3}{*}{D} & Type.5 & \multirow{3}{*}{15} & 15--0\\
              & Type.5 &  & 15--0\\
              & Type.1 &  & 14--4\\\hline
            \multirow{2}{*}{D} & Type.4 & \multirow{2}{*}{6} & 12--0\\
              & Type.3 &  & 36--0\\\hline
            \multirow{2}{*}{D} & Type.4 & \multirow{2}{*}{4} & 18--0\\
              & Type.2 &  & 5--25\\\hline
        \end{tabular}
    \end{subtable}\hfill
    \begin{subtable}[t]{.325\linewidth}
        \centering
        \caption{IELTS / \\FT $\equiv$ E2LLM.}
        \begin{tabular}{|c|c|c|c|}
            \hline
            R & Type & Sl. & G--C\\\hline
            \multirow{2}{*}{P} & Type.4 & \multirow{2}{*}{2} & 10--0\\
              & Type.3 &  & 38--0\\\hline
            P & Type.3 & 1 & 48--0\\\hline
            \multirow{3}{*}{D} & Type.5 & \multirow{3}{*}{15} & 15--0\\
              & Type.5 &  & 15--0\\
              & Type.1 &  & 14--4\\\hline
            \multirow{2}{*}{D} & Type.4 & \multirow{2}{*}{4} & 18--0\\
              & Type.2 &  & 5--25\\\hline
        \end{tabular}
    \end{subtable}\hfill
    \begin{subtable}[t]{.325\linewidth}
        \centering
        \caption{IELTS / Splitwise.}
        \begin{tabular}{|c|c|c|c|}
            \hline
            R & Type & Sl. & G--C\\\hline
            P & Type.1 & 1 & 25--23\\\hline
            D & Type.3 & 2 & 48--0\\\hline
            \multirow{6}{*}{D} & Type.2 & \multirow{6}{*}{18} & 3--3\\
              & Type.4 &  & 3--0\\
              & Type.4 &  & 3--0\\
              & Type.3 &  & 10--0\\
              & Type.5 &  & 13--0\\
              & Type.5 &  & 13--0\\\hline
        \end{tabular}
    \end{subtable}

    \vspace{0.1cm}
    \begin{subtable}[t]{.325\linewidth}
        \centering
        \caption{LZByteDance / \textsc{Spade}.}
        \begin{tabular}{|c|c|c|c|}
            \hline
            R & Type & Sl. & G--C\\\hline
            \multirow{2}{*}{P} & Type.2 & \multirow{2}{*}{2} & 8--20\\
              & Type.4 &  & 20--0\\\hline
            \multirow{3}{*}{D} & Type.5 & \multirow{3}{*}{15} & 16--0\\
              & Type.5 &  & 16--0\\
              & Type.1 &  & 15--1\\\hline
            \multirow{3}{*}{D} & Type.3 & \multirow{3}{*}{12} & 22--0\\
              & Type.4 &  & 7--0\\
              & Type.3 &  & 19--0\\\hline
        \end{tabular}
    \end{subtable}\hfill
    \begin{subtable}[t]{.325\linewidth}
        \centering
        \caption{LZByteDance / \\FT $\equiv$ E2LLM.}
        \begin{tabular}{|c|c|c|c|}
            \hline
            R & Type & Sl. & G--C\\\hline
            \multirow{2}{*}{P} & Type.4 & \multirow{2}{*}{2} & 9--0\\
              & Type.3 &  & 39--0\\\hline
            P & Type.3 & 1 & 48--0\\\hline
            \multirow{3}{*}{D} & Type.5 & \multirow{3}{*}{15} & 16--0\\
              & Type.5 &  & 16--0\\
              & Type.1 &  & 15--1\\\hline
            \multirow{2}{*}{D} & Type.4 & \multirow{2}{*}{4} & 19--0\\
              & Type.2 &  & 5--24\\\hline
        \end{tabular}
    \end{subtable}\hfill
    \begin{subtable}[t]{.325\linewidth}
        \centering
        \caption{LZByteDance / Splitwise.}
        \begin{tabular}{|c|c|c|c|}
            \hline
            R & Type & Sl. & G--C\\\hline
            P & Type.1 & 1 & 25--23\\\hline
            D & Type.3 & 2 & 48--0\\\hline
            \multirow{6}{*}{D} & Type.2 & \multirow{6}{*}{18} & 3--3\\
              & Type.4 &  & 3--0\\
              & Type.4 &  & 3--0\\
              & Type.3 &  & 10--0\\
              & Type.5 &  & 13--0\\
              & Type.5 &  & 13--0\\\hline
        \end{tabular}
    \end{subtable}
    \vspace{-0.5cm}
\end{table}

\paragraph{Workload change} Holding the model at Qwen3MoE, the fleet at eight devices, and the arrival period at 100\,ms fixed, we switch the workload from IELTS to LZByteDance (see Table \ref{tab:dep-workload}). This changes the Prefill-to-Decode token ratio from 1.17 to 0.77, crossing the point at which Decode becomes more demanding than Prefill. A placement strategy that fixes the phase split cannot naturally adapt to this shift; our congestion objective does. On IELTS, \textsc{Spade} uses a single Apple M2 Max for Prefill and dedicates 25 slots to Decode. On LZByteDance, it reconfigures the deployment by increasing the Decode pool to 27 slots. Hence, \textsc{Spade} trades excess Prefill capacity for greater Decode concurrency. Although the per-request decode rate is lower than under FT (28.9 vs 31.3 tokens/s), the larger pool raises aggregate decode throughput by about 31\% (27 vs 19 slots), thus reducing congestion.

In contrast, FT and E2LLM objectives retain the same three Prefill devices across both workloads and \textit{adjust only the layer partition boundaries}, by a single layer.  E2LLM behaves similarly because its objective still includes the Prefill service-time term. Although the workload changes substantially, the reduction in Prefill service time remains large enough to outweigh the benefit of reallocating devices across phases, so the optimizer never leaves the original placement.
The workload change nevertheless affects the underlying execution characteristics. Interestingly, PS reduces whereas DS increases when changing from IELTS to LZByteDance dataset. Because Transformer inference relies on self-attention, the cost of a layer grows with the number of positions already stored in the KV cache. IELTS and LZByteDance therefore induce different execution depths, reflected in their different input and output token counts, and consequently different stage time. The optimal balance between Prefill and Decode is therefore not the same across the two workloads.%, leading to \textit{adjust only the layer partition boundaries}.

Interestingly, FT and E2LLM still converge to the same deployment. Because the search space is discrete, changing a placement requires replacing an entire replica composition rather than making an incremental adjustment. The workload shift is sufficient to alter the objective values, but not sufficient to move the solution into a different basin of the search space. Consequently, FT and E2LLM select the same plan and differ only in the value they assign to it. Adapted Splitwise is even more rigid, keeping the RTX5070 in Prefill for both workloads because its rule always assigns the highest-throughput replica to Prefill regardless of workload characteristics.

\paragraph{Arrival rate change} 

Holding the model, workload, and fleet fixed at Qwen3MoE, LZByteDance, and eight devices, respectively, we vary the arrival period from 100\,ms to 3000\,ms (Table~\ref{tab:dep-arrival}). This is the only configuration that transitions from overload to under-load, with the crossover occurring between 1\,s and 2\,s. Within the overloaded regime, placements remain unchanged: \textsc{Spade} and E2LLM retain their deployments up to 1000\,ms and converge to the same placement at 3000\,ms, whereas finishing time and adapted Splitwise keep the same deployment throughout. The throughput gains remain visible after the transition. At 3000\,ms, the per-request decode rate reaches 59.0 tokens/s under \textsc{Spade} and E2LLM, compared with 46.8 tokens/s under finishing time and 33.1 tokens/s under adapted Splitwise. Once excess Prefill capacity no longer reduces congestion, \textsc{Spade} reallocates resources toward Decode. Combined with its higher decode throughput and larger decode pool at saturation, this allows idle capacity to be converted into higher per-request throughput even after waiting time reaches zero.
The most interesting observation is that \textsc{Spade} and E2LLM eventually converge to the same deployment. Once the offered load falls sufficiently below capacity, additional Prefill resources no longer reduce queueing delay, and both objectives therefore shift resources toward Decode despite being derived from different formulations. 

One exception appears at an arrival period of 3000\,ms, where \textsc{Spade} still exhibits approximately 7\,s of maximum waiting time (Table \ref{tab:objective}). The log from experiments showed that measured Prefill throughput was up to 30\% lower than predicted, causing residual congestion. 
%The discrepancy arises from representing the workload by its average input length: throughput at the average sequence length is not equal to the average throughput across requests because execution cost is a nonlinear function of sequence length. %Consequently, Prefill capacity is slightly overestimated when request lengths vary substantially, causing small bottleneck on Prefill phase. 
The discrepancy arises because the planner represents a workload using a representative sequence length rather than the full request-length distribution. Since latency scales nonlinearly with sequence length, the throughput evaluated at the representative length generally differs from the average throughput over the workload. This approximation substantially reduces planning complexity and is consistent with the workload abstractions commonly used in prior deployment planners \cite{patel2024splitwise,jiang2025thunderserve,jiang2026hexgen3}.
Importantly, this effect is confined to Prefill. Decode throughput remains accurately predicted. The same over-prediction exists in the overloaded configurations, but there the predicted bottleneck matches the measured one, so it changes only the magnitude of congestion. At 3000\,ms the model predicts $W = 0$, yet measurements show residual Prefill congestion.
We also observe small differences in Prefill throughput for SPW despite the same Prefill. These differences are likely caused by device power management under varying idle times. We do not analyze this further, as it concerns runtime behavior rather than placement.

\paragraph{Scalability} We evaluate nested rosters of four, six, and eight devices so that fleet size is never confounded with fleet composition.
The four-device fleet consists of one Type.1, one Type.2, and two Type.4 devices; the 6-device fleet additionally includes one Type.3 and one Type.5 device; and the 8-device fleet uses all devices. 
%The nesting begins from the weaker subset, making the axis a direct measure of the benefit provided by each additional device.
%Table~\ref{tab:dep-fleet} reports the deployments for the four- and six-device fleets, while the eight-device deployments appear in Table~\ref{tab:dep-workload}. 
Across all fleet sizes, \textsc{Spade} consistently achieves the lowest maximum waiting time. The advantage is most pronounced on the smallest fleet, where the finishing-time and E2LLM objectives incur 17\% longer maximum waiting times and adapted Splitwise incurs 160\% longer maximum waiting times relative to \textsc{Spade}. 
%As the fleet grows, these gaps narrow because additional devices provide more placement alternatives and make resource mismatches easier to absorb. 
%Nevertheless, \textsc{Spade} remains the best-performing objective across all three fleet sizes.

\subsection{Framework comparison} \label{subsec:eval:framework}
\textsc{Spade} is built on \texttt{llama.cpp} but replaces the original star-topology communication with a peer-to-peer pipeline. We therefore compare against both the original \texttt{llama.cpp} implementation and E2LLM, which also adopts a star topology. With three devices and nine slots (Type.1, Type.5, Type.2), \textsc{Spade} achieves 37.13 tokens/s, compared with 25.23 tokens/s for E2LLM and 12.69 tokens/s for \texttt{llama.cpp}. At this scale, E2LLM benefits from distributed pipeline execution, whereas the original \texttt{llama.cpp} cannot perform networked pipeline parallelism. The picture changes as the pipeline grows. With six devices and 18 slots (first decoder replica, Table \ref{tab:dep-arrival}.d), \textsc{Spade} sustains 23.92 tokens/s, whereas E2LLM and \texttt{llama.cpp} achieve only 5.02 and 6.29 tokens/s, respectively. The reason is communication contention. In E2LLM, every stage exchanges activations through a central node, forcing multiple pipeline stages to compete for the same uplink and downlink bandwidth. As additional stages are added, communication becomes the dominant bottleneck and eventually outweighs the benefit of pipeline parallelism itself, causing E2LLM to perform even worse than the original \texttt{llama.cpp}. In contrast, the P2P ring distributes traffic across devices and avoids concentrating on a single node, allowing throughput to scale more gracefully with pipeline depth.

\begin{table}[t]
    \centering
    \setlength{\tabcolsep}{2pt}
    %\caption{Deployment plans when varying the arrival rate (eight devices, Qwen3-MoE, LZByteDance).}
    \caption{Deployment plans when varying the arrival rate.}
    \vspace{-0.2cm}
    \label{tab:dep-arrival}
    \begin{subtable}[t]{.48\linewidth}
        \centering
        \caption{\textsc{Spade}, $100$ and $1000$~ms.}
        \begin{tabular}{|c|c|c|c|}
            \hline
            R & Type & Sl. & G--C\\\hline
            \multirow{2}{*}{P} & Type.2 & \multirow{2}{*}{2} & 8--20\\
              & Type.4 &  & 20--0\\\hline
            \multirow{3}{*}{D} & Type.5 & \multirow{3}{*}{15} & 16--0\\
              & Type.5 &  & 16--0\\
              & Type.1 &  & 15--1\\\hline
            \multirow{3}{*}{D} & Type.3 & \multirow{3}{*}{12} & 22--0\\
              & Type.4 &  & 7--0\\
              & Type.3 &  & 19--0\\\hline
        \end{tabular}
    \end{subtable}\hfill
    \begin{subtable}[t]{.48\linewidth}
        \centering
        \caption{\textsc{Spade} $\equiv$ E2LLM, $3000$~ms.}
        \begin{tabular}{|c|c|c|c|}
            \hline
            R & Type & Sl. & G--C\\\hline
            P & Type.2 & 1 & 8--40\\\hline
            \multirow{3}{*}{D} & Type.5 & \multirow{3}{*}{15} & 16--0\\
              & Type.5 &  & 16--0\\
              & Type.1 &  & 15--1\\\hline
            \multirow{2}{*}{D} & Type.4 & \multirow{2}{*}{6} & 12--0\\
              & Type.3 &  & 36--0\\\hline
            \multirow{2}{*}{D} & Type.4 & \multirow{2}{*}{6} & 12--0\\
              & Type.3 &  & 36--0\\\hline
        \end{tabular}
    \end{subtable}

    \vspace{0.1cm}
    \begin{subtable}[t]{.48\linewidth}
        \centering
        \caption{FT, all intervals ($\equiv$ E2LLM at $100$ and $1000$~ms).}
        \begin{tabular}{|c|c|c|c|}
            \hline
            R & Type & Sl. & G--C\\\hline
            \multirow{2}{*}{P} & Type.4 & \multirow{2}{*}{2} & 9--0\\
              & Type.3 &  & 39--0\\\hline
            P & Type.3 & 1 & 48--0\\\hline
            \multirow{3}{*}{D} & Type.5 & \multirow{3}{*}{15} & 16--0\\
              & Type.5 &  & 16--0\\
              & Type.1 &  & 15--1\\\hline
            \multirow{2}{*}{D} & Type.4 & \multirow{2}{*}{4} & 19--0\\
              & Type.2 &  & 5--24\\\hline
        \end{tabular}
    \end{subtable}\hfill
    \begin{subtable}[t]{.48\linewidth}
        \centering
        \caption{Splitwise, all intervals.}
        \begin{tabular}{|c|c|c|c|}
            \hline
            R & Type & Sl. & G--C\\\hline
            P & Type.1 & 1 & 25--23\\\hline
            \multirow{6}{*}{D} & Type.2 & \multirow{6}{*}{18} & 3--3\\
              & Type.4 &  & 3--0\\
              & Type.4 &  & 3--0\\
              & Type.3 &  & 10--0\\
              & Type.5 &  & 13--0\\
              & Type.5 &  & 13--0\\\hline
            D & Type.3 & 2 & 48--0\\\hline
        \end{tabular}
    \end{subtable}
    \vspace{-0.2cm}
\end{table}

\subsection{Limitations of \textsc{Spade}} \label{subsec:eval:limitations}
As mentioned, \textsc{Spade} predicts Prefill throughput accurately for individual operating points, but the throughput of an average request is not necessarily equal to the average throughput across requests. Consequently, workloads represented only by average input lengths can introduce high prediction error when request lengths exhibit high variance. 
%This effect appears in the 3000,ms configuration, where the measured Prefill throughput is approximately 30\% lower than predicted. The error stems from workload aggregation rather than from the per-replica performance model itself.
In addition, \textsc{Spade} targets static deployment planning. The optimizer assumes a fixed workload characterization, arrival rate, and fleet composition during planning. % and does not address runtime orchestration under continuously changing traffic conditions. 
Dynamic adaptations such as online role reassignment, replica migration, or re-optimization under changing arrival rates and request sizes are complementary to our work and remain future directions. Finally, the extension of dominance to descendant replicas is supported empirically rather than proven lossless in general. Exhaustive ones produce identical survivor sets, Pareto frontiers, and final placements wherever verification remains computationally affordable, but a general proof remains open.

\section{Conclusion}\label{sec.conclusion}
We presented \textsc{Spade}, a tractable and scalable placement planner for heterogeneous edge fleets. \textsc{Spade} models disaggregated Prefill/Decode serving as a queueing system and optimizes congestion directly. To make the optimization tractable, it combines pruning with a water-filling formulation for layer partitioning, pruning up to 99.6\% of candidates while preserving the same survivor set as exhaustive ones. Combined with water filling, \textsc{Spade} reaches the verified optimum in hundreds of milliseconds on a 30-device fleet, whereas heuristic searches remain substantially below optimum even after seconds to hours of execution. These results suggest that optimal queueing-aware placement for heterogeneous edge serving is tractable at fleet scale.

%\begin{comment}
\section*{Acknowledgment}
This work was supported in part by the Norwegian Research Council under Grant 322473 (AirQMan project) and the European Commission under Grant 101086541 (MISO project).
%\end{comment}

\printbibliography

\end{document}